\documentclass[aps,prd,reprint,nofootinbib,twocolumn]{revtex4-2}
\pdfoutput=1
\usepackage[english]{babel}
\usepackage{appendix}
\usepackage{amsmath}
\usepackage{amsfonts}
\usepackage{graphicx}
\usepackage[colorlinks=true, allcolors=blue]{hyperref}
\usepackage{cancel}
\def\L3{\Lambda_3}

\newcommand{\VSI}{Van Swinderen Institute for Particle Physics and Gravity,\\ University of Groningen,
Nijenborgh 4, 9747 AG Groningen, The Netherlands}
\newcommand{\SNS}{Scuola Normale Superiore and INFN, Piazza dei Cavalieri 7, 56126, Pisa, Italy}

\begin{document}
\title{Non-linear Electrodynamics from Massive Gravity}
\author{Thomas Fl\"oss}
\email{t.s.floss@rug.nl}
\author{Diederik Roest}
\email{d.roest@rug.nl}
\affiliation{\VSI}
\author{Tom Westerdijk}
\email{tom.westerdijk@sns.it}
\affiliation{\SNS}

\begin{abstract}
\noindent
As a counterpart to the four-fermion interaction, which describes massive vector exchange at low energies, we investigate the low energy effective action of photons under exchange of a massive graviton. We show how integrating out a massive graviton leads to the most general duality-invariant vector interactions in 4D or, vice versa, how any such interactions have a natural interpretation within massive gravity. Moreover, we demonstrate how the special case of Born-Infeld theory arises from arguably the simplest graviton potential within ghost-free dRGT massive gravity.
\end{abstract}

\maketitle

\section{Introduction}

\noindent
Effective field theories are ubiquitous throughout physics. Perhaps the most famous example is the low-energy description of $\beta$-decay in terms of four-fermion interactions \cite{Fermi:1934hr}. With the development of the electroweak theory and the Higgs mechanism \cite{Higgs:1964pj,Englert:1964et}, it became clear that at a more fundamental level this physical process is mediated by the $W_{\pm}$ and $Z_0$ vector bosons. Fermi's theory then arises as a low energy description of this process, valid below the electroweak scale of $\sim100$ GeV, in terms of effective contact interactions for the fermions. 

This top-down approach to effective field theories can be more broadly applied to any theory with a separation of scales between lighter and heavier degrees of freedom: at energies well below the heavy mass, there is insufficient energy available to excite the heavier modes. These can therefore be integrated out,  resulting in effective interactions between the lighter degrees of freedom. This philosophy is independent of the nature of the heavier degrees of freedom and can also be applied to massive gravitons.

Massive gravity (MG), originally formulated with an eye towards the cosmological constant problem, is a highly non-trivial and interesting construction from a purely theoretical perspective. Remarkably, ensuring the absence of ghosts allows for two free parameters (besides the graviton mass) in 4D \cite{deRham:2010ik,deRham:2010kj}; in contrast to the electroweak theory, no Higgs-like mechanism is known to connect the massive and massless phases. Moreover, the parameter space of MG is subject to positivity bounds \cite{Cheung:2016yqr}, with a very recent claim that these do not allow for a parametric separation of graviton mass well below the cut-off scale \cite{Bellazzini:2023nqj}. 

In this paper, we will consider effective theories emerging from integrating out massive gravitons (and thus the opposite limit of \cite{Bellazzini:2023nqj}). For simplicity we will take the lighter degrees of freedom to be massless, and focus on the case of a Maxwell vector that only interacts gravitationally. The resulting low-energy description, with effective self-interactions of the Maxwell vector, spans an interesting set of theories: it consists of all non-linear completions that preserve the electromagnetic duality invariance of the original Maxwell theory\footnote{Duality invariance also plays an important role in supersymmetric theories, where e.g.~the $E_{7(7)}$ duality invariance features prominently in the discussion of possible finiteness of $\mathcal{N}=8$ supergravity \cite{Beisert:2010jx,Bossard:2010dq}.}  (in contrast to e.g.~the Euler-Heisenberg theory \cite{Heisenberg:1936nmg} that follows from integrating out massive electrons at one-loop). All of these effective theories are expected to be fully compatible with positivity bounds; see Ref.~\cite{Henriksson:2021ymi, CarrilloGonzalez:2023cbf} for such bounds for $2\rightarrow2$ scattering. 

In the second half of this paper, we will restrict ourselves to the ghost-free dRGT theory of massive gravity \cite{deRham:2010ik, deRham:2010kj}. We demonstrate how, starting from a particularly simple element of dRGT, one can generate the most important example of non-linear electrodynamics: Born-Infeld theory. This theory describes the world-volume of D-branes in string theory \cite{Polchinski:1995mt} and has numerous special amplitude properties, see e.g.~\cite{Cachazo:2014xea, Cheung:2018oki}. It is therefore interesting that BI arises as an effective description of graviton exchange. Moreover, Born-Infeld and gravity-coupled Maxwell theories both appear in the double copy framework of \cite{Bern:2010ue}, as illustrated in the tetrahedron of \cite{deNeeling:2022tsu}; it thus appears that moving to massive exchange particles corresponds to a particular operation in the double copy web of relations, similar to \cite{Cachazo:2014xea, Cheung:2017ems}.   

We comment on further links and implications and future extensions in the concluding section. Amongst these is the analogon of our discussion in 2D space-time, where the idea of integrating out massive gravity was worked out in 2D in Ref.~\cite{Tolley:2019nmm}. It was shown that coupling dRGT gravity to a generic field theory is equivalent to performing a $T \bar{T}$-deformation of said theory. Our work can thus be seen as the extension of this procedure to the case of conformally-coupled Maxwell theory in 4D, suggesting a close connection to $T\bar{T}$-deformations of this theory, as studied in Refs.~\cite{Cavaglia:2016oda,Conti:2018jho,Babaei-Aghbolagh:2022uij,Ferko:2022iru,Conti:2022egv}.

\noindent
\textbf{Notation.} We work in 4D spacetime throughout and employ matrix notation, with e.g.~$F \equiv F_{\mu \nu}$ and  $F^2 \equiv  F_{\alpha \gamma} \eta^{\gamma \delta} F_{\delta \beta}$ (note the ordering of indices). Traces taken with the flat metric $\eta$ are written using square brackets, e.g. $[F^2] = F_{\alpha \beta} \eta^{\beta \gamma} F_{\gamma \delta} \eta^{\delta \alpha}$. Angled brackets instead denote traces where all tensors are contracted with the full metric $g$, e.g. $\langle F^2 \rangle = F_{\alpha \beta} g^{\beta \gamma} F_{\gamma \delta} g^{\delta \alpha}$ (following the conventions of \cite{Hinterbichler:2011tt}). 

\section{Integrating out gravity}
\label{sec:Pert}
\noindent
We start by integrating out the massive graviton perturbatively, obtaining the effective vector contact interactions order by order. Our set-up consists of a Maxwell field minimally coupled to massive gravity: 
\begin{align}
\label{eq:gMax}
    \mathcal{L}_{gF^2} + \mathcal{L}_g = \sqrt{-g} \left[ \tfrac14 \langle F^2 \rangle  + \tfrac12 R  - \tfrac18 m^2 \sum_{i} V^{(i)}(h)  \right] \,.
\end{align}
where $g = \eta + h$. The most general graviton potential terms are given at lowest order by \cite{Hinterbichler:2011tt}
  \begin{align}
     V^{(2)}  = & \langle h^2\rangle + b_2 \langle h\rangle^2 \,, \notag \\
     V^{(3)}  = & c_1 \langle h^3\rangle + c_2 \langle h\rangle \langle h^2\rangle + c_3 \langle h\rangle^3   \,, \notag \\
     V^{(4)}  = & d_1 \langle h^4\rangle + d_2 \langle h^3\rangle\langle h\rangle + d_3 \langle h^2\rangle^2 + d_4 \langle h^2\rangle\langle h\rangle^2 + \notag  \\ 
    & +  d_5 \langle h\rangle^4 \,.
 \end{align}
At energies well below the mass scale of the graviton, we can neglect the Einstein-Hilbert term so the massive graviton becomes an auxiliary field and can be integrated out through its algebraic equation of motion\footnote{Theories of massive gravity typically suffer from ghosts, for which it is inconsistent to neglect the Einstein-Hilbert term. We will return to this issue in section \ref{sec:L3}.},
\begin{equation}
    \frac{1}{\sqrt{-g}}\frac{\delta\sqrt{-g}V(g)}{\delta g^{-1}} = T^{(\text{Maxwell})} \,,
    \label{eq:constraints}
\end{equation} 
in matrix notation. 

Proceeding perturbatively, we expand the metric as $h = h^{(1)} + h^{(2)} + \ldots$ and solve the above equation of motion order by order, resulting in effective contact interactions for the vector field.
At lowest order, the graviton field equation is solved by
\begin{align}
    h^{(1)} =  - \frac{2}{m^2} F^2 + \frac{1}{2m^2} [F^2] \eta \,.
\end{align}
When plugged back into the action, it gives rise to the following effective four-point contact interactions (setting $m^2=1$ for brevity):
 \begin{align}
  \mathcal{L}_{F^4} = \tfrac{1}{2} [F^4] - \tfrac{1}{8}[F^2]^2\,.
 \end{align}
Note that $b_2$ drops out at this order due to the tracelessness of the lowest-order solution. Moving to the next order, the solution reads
\begin{align}
   h^{(2)} & = -2[F^2] F^2 + f_1 [F^4] \eta + f_2 [F^2]^2 \eta \,,
\end{align}
for functions $f_{1,2}$ of the parameters $(b_2,c_1,c_2)$ whose specific forms are not relevant for our purposes. We have rewritten an $F^4$ term using the following recursive matrix identity: 
\begin{align}
\label{eq:matrixrel}
    X^n & = \tfrac12 X^{n-1}[X] + \tfrac14 X^{n-2} [X^2] - \tfrac18 X^{n-2}[X]^2, \,\, n \geq 2 \,, \notag \\
\end{align}
for $X=F^2$. Moreover, it is also valid for $X=h$ due to the fact that $h$ is solved for in terms of $F^2$. Upon plugging the solution back into the action, we obtain the six-point contact interactions:
\begin{eqnarray}
    \mathcal{L}_{F^6} =  \tfrac{1}{2}[F^4][F^2] -\tfrac{1}{8}[F^2]^3 \,.
\end{eqnarray}
Remarkably, all MG coefficients again drop out at this order; ultimately this can again be traced back to the tracelessness of the lowest-order solution $h^{(1)}$, such that the new coefficients only appear in the trace part of $h^{(2)}$ - which drops out at the $F^6$ order. 

The parametric freedom in the MG potential therefore only comes in at octic order: the coefficients of the three terms $[F^4]^2$, $[F^4] [F^2]^2$ and $[F^2]^4$ will depend on the first five parameters $(b_2,c_1,c_2,d_1,d_3)$. Moreover, the ten-point interactions are fixed in terms of the eight-point interactions and hence depend on the same combination of these five parameters. Once we move to the twelve- and fourteen-point interactions, four new combinations of nine additional parameters of the graviton potential appear. 

We therefore find that massive gravity leads to a very specific non-linear modification of Maxwell's free theory: the massive gravity potential induces a new interaction at every order in $F^4$, with corresponding  terms at higher orders in $F^2$. As we will see below, these are precisely those theories of non-linear electrodynamics that are invariant under electromagnetic duality.

\section{Duality-invariant non-linear electrodynamics}\label{sec:duality}

\noindent
Maxwell's equations for a free photon are famously invariant under transformations that rotate the electric and magnetic fields into each other:
\begin{align}
    F_{\mu \nu} \rightarrow \cos{\alpha}~ F_{\mu \nu} + \sin{\alpha}~\tilde{F}_{\mu \nu} \,.
\end{align}
These transformations are often referred to as 'duality' rotations between the field strength and its Hodge dual $\tilde{F}_{\mu \nu} \equiv \frac{1}{2}\varepsilon_{\mu \nu \rho \sigma} F^{\rho \sigma}$. Given the central role of symmetries in high energy physics, a natural question is whether there are non-linear extensions of Maxwell theory that preserve this duality invariance.

The most general theory of this type was constructed by Gibbons and Rasheed\footnote{A somewhat similar construction was already presented in \cite{Bialynicki-Birula:1984daz}. Further details on the history of this theory can be found in \cite{Sorokin:2021tge}.} \cite{Gibbons:1995cv} and is most conveniently written in terms of two invariants 
 \begin{align}
   p_{\pm} = \tfrac{1}{8}\left([F^2] \pm \sqrt{4 [F^4] - [F^2]^2 }\right) \,.
 \end{align}
Duality invariance imposes the following constraint on the Lagrangian:
 \begin{align}
 \label{eq:duality-requirement}
  \mathcal{L}_- \mathcal{L}_+ = 1 \,,
 \end{align}
where $\mathcal{L}_\pm$ denotes the derivative of $\mathcal{L}$ with respect to $p_\pm$. The most general solution to this requirement was subsequently given by Gaillard and Zumino \cite{Gaillard:1997rt} and involves the freedom to specify an arbitrary function $v(s)$. Its argument is solved for in terms of $p_\pm$ via
  \begin{align}
  \label{eq:s-constraint}
      p_+ = v'^2 (p_- -s) \,,
    \end{align}
after which the Lagrangian takes the form
 \begin{align}
 \label{eq:hilbert}
     \mathcal{L} = \frac{2 p_+}{v'} + v \,.
 \end{align}
There are two well-known closed-form solutions to \eqref{eq:s-constraint}. As a simple generalisation of Maxwell ($v=s$), one can take $v=s \exp(c)$ for some constant $c$. This leads to the so-called ModMax theory,
 \begin{align}
  \mathcal{L} = \tfrac14 \cosh(c) [F^2] + \tfrac14 \sinh(c) \sqrt{4 [F^4] - [F^2]^2}
 \end{align}
with the special property of being the only modification that, on top of duality invariance, also preserves conformal symmetry \cite{Bandos:2020jsw}. This Lagrangian is not analytic when expanded around an $F=0$ vacuum, however, and hence can only be used for non-trivial electromagnetic backgrounds. To ensure analyticity around $F=0$, the Lagrangian has to be even under the interchange of $p_\pm$; we will restrict to this class from now onwards. The second closed form solution, with analyticity, is Born-Infeld theory \cite{Born:1934gh}, to which we will turn in the next section. Further generalisations were found in~\cite{Hatsuda:1999ys}. 

In order to connect to massive gravity, one should first note that the above system of equations \eqref{eq:s-constraint}, \eqref{eq:hilbert} allows for an interpretation of $s$ as an auxiliary field: at the solution for  $s$, the Lagrangian can be written as 
 \begin{align}
 \label{eq:GZaux}
      \mathcal{L} = \frac{1}{v'} p_+ + v' p_-  - V_s \,, \quad V_s = s v' - v \,.
 \end{align}
Provided $v'' \neq 0$, the constraint \eqref{eq:s-constraint} is then actually equivalent to the field equation for $s$. In this new formulation, the duality requirement \eqref{eq:duality-requirement} is moreover manifestly satisfied. It thus follows that the construction of the most general non-linear duality-invariant electrodynamics corresponds to integrating out an auxiliary field\footnote{This approach can be extended to supersymmetric systems, see e.g.~\cite{Kuzenko:2000tg, Ivanov:2013ppa, Carrasco:2011jv}.} \cite{Hatsuda:1999ys, Ivanov:2002ab}; however, what remains to be explained are the very specific couplings $v'^{\pm 1}$ of $s$ to the two components $p_\pm$ of the Maxwell field. 

Remarkably, these follow exactly from integrating out massive gravity, and thus are determined by diffeomorphism invariance and minimal coupling. To show this, we will proceed in three steps. First observe that solving the metric, $g$, in terms of the field strength, $F^2$, and a background metric $\eta$ via the constraint equation \eqref{eq:constraints} implies that the matrices corresponding to $g$ and $F^2$ must commute. In other words, the two are simultaneously diagonalizable and, as a consequence, the eigenvalues of $g$ come in two pairs with distinct values, i.e.~$\lambda_{1,2}$. 
This confirms the intuition one obtains from the perturbative approach of the previous section, where the metric is derived in terms of a polynomial in the field strength; its most general, non-perturbative expression is 
\begin{equation}
    g = c_1(g,F) \eta + c_2(g,F) F^2,
\end{equation}
where $c_{1,2}$ are scalar functions of the metric and field-strength traces. 
We can fix these in terms of traces as
\begin{align}
  g = \tfrac14 [g] \eta + \sqrt{\frac{[g]^2-4[g^2]}{[F^2]^2-4[F^4]}} \left( F^2 - \tfrac{1}{4} [F^2] \eta \right), \label{ginF}
\end{align}
to ensure that the metric obeys the correct relation to the eigenvalues $[g^i] = 2 (\lambda_1^i + \lambda_2^i)$. 

Secondly, since the Maxwell term in the Lagrangian is conformally invariant, it can only depend on the ratio of these two eigenvalues. The specific expression follows from plugging \eqref{ginF} into \eqref{eq:gMax} and turns out to be particularly simple:
\begin{equation}
   \mathcal{L}_{gF^2} = c_+ p_+ + c_- p_-  \,, \quad c_\pm = (\lambda_1 / \lambda_2)^{\pm 1} \,.
    \label{eq:MGF2}
\end{equation}
Note the close similarity to the auxiliary field formulation \eqref{eq:GZaux}; however, we have now {\it derived}, instead of postulated, the specific couplings to $p_\pm$ from massive gravity.

Finally, in contrast to the Maxwell part, the graviton potential in equation \eqref{eq:gMax} is not conformally invariant and therefore depends more generally on the two eigenvalues rather than their ratio only. A convenient approach is to separate off a conformal factor, $g = \rho^2 \tilde g$. This factor contributes via the MG potential to the trace (taken with the inverse metric) of the graviton field equation \eqref{eq:constraints}. Solving for it via this trace equation (or, alternatively, extremising with respect to the conformal factor itself) yields a 'reduced' MG potential with conformal invariance, which only depends on the ratio of eigenvalues (of either $g$ or $\tilde g$). Moreover, its dependence has to be invariant under relabelling of the eigenvalues. The simplest quantity with this property is: 
\begin{eqnarray}
\label{eq:f2} 
    f^2 \equiv \frac{\lambda_1}{\lambda_2} + \frac{\lambda_2}{\lambda_1} - 2  \,.
\end{eqnarray}
Crucially, this is small  if the auxiliary metric is close to flat-space, with the first contributions coming in at quadratic order: $f^2 = [h^2]- \tfrac14[h]^2 + \mathcal{O}(h^3)$; note that odd powers of $f$ are not analytic around $h=0$. Without loss of generality, the most general EFT can thus be generated by a reduced MG potential,
 \begin{align}
 \label{eq:scalpot}
  V_f = \sum_{n} a_n f^{2n} \,,
 \end{align}
where the coefficients, $a_i$, are generated by the original MG potential in the process of integrating out $\rho$. Expressing \eqref{eq:MGF2} in the same combination yields
 \begin{align}
 \label{eq:cpm}
  c_\pm = 1 + \tfrac12 f^2 \pm \tfrac 12 \sqrt{f^2 (4+f^2)} \,.
 \end{align}
which preserves duality invariance according to \eqref{eq:duality-requirement}, since $c_- c_+ = 1$. 

At this point it is simple to identify the mapping between the formulations with auxiliary fields $s$ and $f^2$. For a given $v(s)$, one can relate the two auxiliary fields by $c_- = v'$, which can be plugged into the potential to determine to which $V_f$ this corresponds. Conversely, for a given $V_f$, the same two equations allow one to solve for $v'$ in terms of $v$ and $s$, which can then be solved. There is thus a one-to-one mapping between the auxilary field formulations in terms of $v(s)$ and in terms of $V_f$. We will provide an explicit example of this in the next section. 

\section{dRGT \& Born-Infeld}
\label{sec:L3}
\noindent
So far we have considered the most general graviton potential in equation \eqref{eq:gMax}. However, MG theories typically suffer from negative energy states known as \emph{ghosts}, see e.g. Ref.~\cite{Boulware:1972yco}. In order for our theories to be physical, we therefore have to restrict ourselves to ghost-free theories, requiring a very specific tuning of the coefficients of the general MG potential in equation \eqref{eq:gMax}, known as dRGT gravity \cite{deRham:2010ik,deRham:2010kj}.

At the linearised level this amounts to the Fierz-Pauli tuning $b_2 = -1$ \cite{Fierz:1939ix}, while its non-linear extension is defined in terms of a new tensor quantity \cite{deRham:2010kj}:
\begin{align}
    {\mathcal{K}^\mu}_\nu = {\delta^\mu}_\nu - \sqrt{{\delta^\mu}_\nu -g^{\mu \alpha}h_{\alpha \nu}} \,,
\end{align}
or, in matrix notation, $\mathcal{K} = \mathcal{I} - \sqrt{g^{-1} \eta}$. MG in 4D is then ghost-free for a two parameter family of potentials
\begin{align}
    \mathcal{L}_{\L3} & = -\frac{1}{8}m^2 \sqrt{-g} \sum_{n =  2}^4 \alpha_n \mathcal{L}^{(n)}_{\mathcal{K}} \,, \notag \\
    \mathcal{L}_{\mathcal{K}}^{(2)} &= \left[\mathcal{K}\right]^2 - \left[\mathcal{K}^2\right]\,, \notag \\
    \mathcal{L}_{\mathcal{K}}^{(3)} &= \left[\mathcal{K}\right]^3 - 3 \left[\mathcal{K}\right]\left[\mathcal{K}^2\right] + 2\left[\mathcal{K}^3\right]\,,\notag \\
    \mathcal{L}_{\mathcal{K}}^{(4)} &= \left[\mathcal{K}\right]^4 -6\left[\mathcal{K}^2\right]\left[\mathcal{K}\right]^2+ 8\left[\mathcal{K}^3\right]\left[\mathcal{K}\right] + 3 \left[\mathcal{K}^2\right]^2- 6\left[\mathcal{K}^4\right] \,,
\end{align}
where $\alpha_i = (-4, 2^3 c_3, 2^4 d_5)$ and we will explicitly include the graviton mass $m$ in this section. Another important consequence of this tuning is the fact that the UV-cutoff of the theory is raised from $\Lambda_5 = (M_{\rm P} m^4)^{1/5}$ to $\Lambda_3 = (M_{\rm P}m^2)^{1/3}$. This theory is therefore also referred to as $\L3$-theory. 

We will demonstrate how dRGT gravity is connected to a very specific non-linear completion due to Born and Infeld \cite{Born:1934gh}, which has the remarkable property that it can be formulated geometrically:
 \begin{align}
 \label{eq:BI}
    \mathcal{L}_{\rm BI} = -\tfrac{m^2}{4}\left(\left(-\det(\eta -\tfrac{4}{m^2}F^2)\right)^{1/4} - 1 \right) \,.
\end{align}
It arises in a variety of contexts, including string theory (as the world-volume theory of D-branes \cite{Polchinski:1995mt}), the double copy (allowing for a CHY formulation of its amplitudes \cite{Cachazo:2014xea}) and soft limits (under a multi-chiral soft limit \cite{Cheung:2018oki}). This prompts the question whether it follows from the ghost-free dRGT theory and, if so, what its graviton potential is. 

To this end, we will first  derive the reduced graviton potential corresponding to Born-Infeld theory. Following the example of Ref.~\cite{Gaillard:1997rt}, the auxiliary field formulation of Born-Infeld in terms of $s$ is\footnote{Notice a sign difference compared to Ref.~\cite{Gaillard:1997rt}, which is due to their $p,q$ corresponding to $-p_-,-p_+$ in our notation.}
\begin{align}
    v(s) &= -\frac{m^2}{4} \left(\sqrt{1-\frac{8}{m^2}s}-1 \right) \,.
\label{eq:GaillardBI}
\end{align}
Following the general procedure outlined in the previous section, this can be mapped onto the following auxiliary field formulation in terms of $f^2$:
\begin{align}
\label{eq:VfBI}
    V_f
    = \frac{m^2}{8} \frac{(v'-1)^2}{v'} 
    = \frac{m^2}{8} f^2 \,,
\end{align}
where in the last equality we have made use of the relation \eqref{eq:cpm}. Hence, we conclude that the simplest possible reduced MG potential, that is only linear in $f^2$, leads to Born-Infeld. 

It remains to see whether this fits within dRGT. Following the perturbative approach of section \ref{sec:Pert} and matching order by order, we find that Born-Infeld theory requires the dRGT parameters $c_3 = \tfrac16$ and $d_5=0$. For these values, the full potential can be written as 

\begin{equation}
    \frac{1}{m^2}\mathcal{L}_{\L3}= 1 + \sqrt{-g} \Big([\mathcal{K}]-\det(\mathcal{K})-1\Big) \,,
    \label{eq:WBI}
\end{equation}
where we have used $\det(\mathcal{I}-\mathcal{K}) = \det(g)^{-1/2}$. 
To integrate out the graviton, we follow the procedure described in section \ref{sec:duality}. First, we rewrite the above Lagrangian in terms of the metric, 
\begin{align}
    \frac{1}{m^2}\mathcal{L}_{\L3}= & 1 + \sqrt{-g} \Big( 2  + [g^{-1/2}]^2 ( -\tfrac12  +\tfrac18 [g^{-1/2}]  + \notag \\
    & - \tfrac{1}{64} [g^{-1/2}]^2 ) + [g^{-1}] ( \tfrac12 - \tfrac{1}{4} [g^{-1/2}] + \notag \\ 
    & + \tfrac{1}{16} [g^{-1/2}]^2 - \tfrac{1}{16} [g^{-1}] )     \Big) \,,
    \label{eq:WBI}
\end{align}
where $g^{-1/2} \equiv \sqrt{g^{-1} \eta}$. We then pull out a conformal factor from the metric, $g = \rho^2 \tilde g$ and solve for it using the trace of \eqref{eq:constraints}, or alternatively simply vary with respect to $\rho$ itself. For the specific potential of \eqref{eq:WBI} this yields the particularly simple solution $ \rho = [{\tilde g}^{-1/2} ]$. 
It follows naturally that the resulting reduced MG potential is conformally invariant which allows us to work with the original metric $g$ again. After this substitution, the full Lagrangian can be compactly written as
\begin{eqnarray}
\label{eq:MaxL3}
    \mathcal{L} = \tfrac14 \sqrt{-g}\langle F^2 \rangle - \tfrac{m^2}{16} \sqrt{-g}\langle \eta^2 \rangle +\tfrac{m^2}{4} \,,
\end{eqnarray}
where $\langle \eta^2 \rangle = [g^{-2} ]$ in our notation.
Upon varying and solving \eqref{eq:MaxL3} for $g$, with the simple solution 
\begin{align}
 g = \eta\sqrt{\mathbb{I} - \frac{4}{m^2} F\cdot\eta\cdot F} \,,
\end{align}
this results precisely in Born-Infeld theory with the coupling scale set by the mass of the graviton as given in equation \eqref{eq:BI}.

We thus find that Born-Infeld follows from integrating out the massive graviton in a specific dRGT theory. Moreover, the relevant part of the MG potential in \eqref{eq:MaxL3} indeed takes the simple form \eqref{eq:VfBI} predicted by the auxiliary field prescription. From this perspective, Born-Infeld corresponds to the leading non-linear completion of Maxwell theory, with the simplest possible conformally invariant potential $f^2$. Higher-order terms of $f^2$ in the reduced graviton potential encode the other parameter choices of dRGT.

As a corollary remark, we note that for the choice of parameters $c_3 = \tfrac16$ and $d_5=0$, the helicity-0 component of the theory in the decoupling limit, generally given by the cubic, quartic and quintic Galileon terms, now reduce to the quartic special Galileon (see e.g. equation (9.27) in Ref.~\cite{Hinterbichler:2015pqa}).

\section{Conclusions}

In this work we have studied the low energy effective action of Maxwell theory coupled to massive gravity by integrating out the massive graviton. The invariance of Maxwell theory under duality rotations is retained in this scenario, as gravity is unaffected by such transformations. Hence, we have found a reformulation of the previous classification of duality invariant non-linear electrodynamics \cite{Gibbons:1995cv, Gaillard:1997rt} in terms of an auxiliary graviton field\footnote{Note that auxiliary metrics have also been introduced in the construction of self-dual $p$-form actions \cite{Sen:2019qit, Hull:2023dgp}; it would be interesting to investigate possible connections.}. The conformal invariance of Maxwell plays an important role in this approach, and allows the MG potential to be truncated to a conformal section of it. Remarkably, it turns out that one can generate all non-linear, analytic completions of Maxwell\footnote{Perhaps the same applies to ModMax, such that its non-linear Born-Infeld completion \cite{Bandos:2020hgy} can be derived by coupling it to a massive graviton. We leave this for future work.} in this manner. Although an auxiliary field formulation of non-linear electrodynamics is not new (e.g. \cite{Hatsuda:1999ys, Ivanov:2002ab}), the specific couplings that are needed now follow from  minimal coupling and thus diffeomorphism invariance, instead of being included \emph{ad hoc}. Therefore, our approach naturally preserves principles such as locality and, moreover, has an obvious generalization to include derivative interactions.

Restricting ourselves to the subset of ghost-free massive gravity theories known as dRGT gravity, we identified a particular theory whose conformal MG potential takes a simple quadratic form in terms of the relabeling-invariant combination \eqref{eq:f2} of the graviton's eigenvalues. Integrating out the remaining graviton components leads to the geometric Born-Infeld action to all orders. Given the latter's special properties in terms of amplitudes, it would be interesting to investigate what the current perspective implies. For instance, we have restricted ourselves to the lowest order in the expansion of scattering energy over graviton mass; keeping subleading terms would correspond to higher-derivative terms, e.g.~of the form $\partial^4 F^4$. Can these be formulated in terms of curvature invariants of the metric $\eta + F$? Similarly, do these higher-derivative corrections fit in the CHY formalism \cite{Cachazo:2013hca, Cachazo:2014xea}?

Furthermore, one can wonder whether there is something special about non-linear completions obtained from dRGT massive gravity, since these follow from integrating out a well defined physical graviton field, compared to theories obtained from non-dRGT massive gravity, where one naively integrates out an auxiliary field that propagates ghosts. For example, it would be interesting to see whether the additional dRGT parameters (beyond the specific point with the MG potential \eqref{eq:WBI}) fit in the double copy framework, and moreover whether there is a connection to positivity bounds on these theories \cite{Henriksson:2021ymi, CarrilloGonzalez:2023cbf} as well.

On a different note, given the crucial role of conformal invariance in our approach, it is natural to ask whether a similar construction can be applied to a  scalar field in 2D. In fact, this case is even more special: massive gravity does not propagate any physical degrees of freedom in two dimensions and is therefore a genuine auxiliary field. Moreover, the dRGT graviton potential has no free parameters in 2D. Indeed it has been found that it maps uniquely to the (multi-field) Dirac-Born-Infeld (DBI) theory by eliminating the non-dynamical degrees of freedom~\cite{Tolley:2019nmm}.
One can view the above as the massive extension of the usual transition from the Polyakov to the Nambu-Goto form of the string world-sheet action. In the massless case, the auxiliary world-sheet metric is only determined up to an overall factor, due to the conformal symmetry. The introduction of the dRGT potential in 2D breaks this symmetry and can be seen as a gauge fixing term. Integrating out the world-sheet metric now also fixes its overall factor, but this does not change the resulting expressions: the overall factor drops out of the massless part, and moreover the dRGT potential exactly vanishes for this choice (as it acts as a gauge fixing term for the conformal symmetry).

DBI is special: it describes the longitudinal modes of D-branes and non-linearly realises a higher-dimensional version of Poincar\'e; as a consequence, its amplitudes display a generalised Adler zero in their soft limit \cite{Cheung:2014dqa}. It would be interesting to see whether one can also generate the unique Volkov-Akulov fermionic interactions \cite{Volkov:1973ix} with these properties in this way.

Given the connection between integrating out dRGT gravity and $T\bar{T}$-deformations in 2D \cite{Tolley:2019nmm}, it would be interesting to investigate the possible relation of our approach to $T \bar T$-deformations in 4D, see e.g.~\cite{Cavaglia:2016oda, Conti:2018jho} for Maxwell and \cite{Babaei-Aghbolagh:2022uij, Ferko:2022iru, Conti:2022egv} for ModMax.

In contrast to the special cases discussed here, one loses conformal invariance in generic dimensions. One can easily check (e.g.~by a perturbative approach at low orders) that naively integrating out a massive graviton in such cases does not lead to the specific turning of Born-Infeld for a vector field (or DBI for a scalar field), again underlining the importance of conformal invariance. In order to move into other dimensions and theories, it might be necessary to include dilaton interactions, on top of gravity; this combination is natural from e.g.~the common sector of string theory, as well as the double copy \cite{Bern:2010ue, Cachazo:2013hca}. However, these set-ups are normally massless; our approach would require an extension of massive gravity to also include a dilaton. We leave these considerations for future work.

\section*{Acknowledgements}
\noindent
The authors would like to thank Tom\'{a}\v{s} Brauner, Christian Ferko, Karol Kampf, Sergei Kuzenko, Johannes Lahnsteiner, Gabriele Mazzucchelli, Guilherme Pimentel, Liam Smith and Dmitri Sorokin for useful discussions and insightful comments. TF is supported by the Fundamentals of the Universe research program within the University of Groningen. The Mathematica packages xAct \cite{xAct} and xTras \cite{Nutma:2013zea}
were used extensively in the course of this work.

\bibliographystyle{apsrev4-1}
\bibliography{bibliography.bib}

\begin{thebibliography}{47}%
\makeatletter
\providecommand \@ifxundefined [1]{%
 \@ifx{#1\undefined}
}%
\providecommand \@ifnum [1]{%
 \ifnum #1\expandafter \@firstoftwo
 \else \expandafter \@secondoftwo
 \fi
}%
\providecommand \@ifx [1]{%
 \ifx #1\expandafter \@firstoftwo
 \else \expandafter \@secondoftwo
 \fi
}%
\providecommand \natexlab [1]{#1}%
\providecommand \enquote  [1]{``#1''}%
\providecommand \bibnamefont  [1]{#1}%
\providecommand \bibfnamefont [1]{#1}%
\providecommand \citenamefont [1]{#1}%
\providecommand \href@noop [0]{\@secondoftwo}%
\providecommand \href [0]{\begingroup \@sanitize@url \@href}%
\providecommand \@href[1]{\@@startlink{#1}\@@href}%
\providecommand \@@href[1]{\endgroup#1\@@endlink}%
\providecommand \@sanitize@url [0]{\catcode `\\12\catcode `\$12\catcode
  `\&12\catcode `\#12\catcode `\^12\catcode `\_12\catcode `\%12\relax}%
\providecommand \@@startlink[1]{}%
\providecommand \@@endlink[0]{}%
\providecommand \url  [0]{\begingroup\@sanitize@url \@url }%
\providecommand \@url [1]{\endgroup\@href {#1}{\urlprefix }}%
\providecommand \urlprefix  [0]{URL }%
\providecommand \Eprint [0]{\href }%
\providecommand \doibase [0]{http://dx.doi.org/}%
\providecommand \selectlanguage [0]{\@gobble}%
\providecommand \bibinfo  [0]{\@secondoftwo}%
\providecommand \bibfield  [0]{\@secondoftwo}%
\providecommand \translation [1]{[#1]}%
\providecommand \BibitemOpen [0]{}%
\providecommand \bibitemStop [0]{}%
\providecommand \bibitemNoStop [0]{.\EOS\space}%
\providecommand \EOS [0]{\spacefactor3000\relax}%
\providecommand \BibitemShut  [1]{\csname bibitem#1\endcsname}%
\let\auto@bib@innerbib\@empty
\bibitem [{\citenamefont {Fermi}(1934)}]{Fermi:1934hr}%
  \BibitemOpen
  \bibfield  {author} {\bibinfo {author} {\bibfnamefont {E.}~\bibnamefont
  {Fermi}},\ }\href {\doibase 10.1007/BF01351864} {\bibfield  {journal}
  {\bibinfo  {journal} {Z. Phys.}\ }\textbf {\bibinfo {volume} {88}},\ \bibinfo
  {pages} {161} (\bibinfo {year} {1934})}\BibitemShut {NoStop}%
\bibitem [{\citenamefont {Higgs}(1964)}]{Higgs:1964pj}%
  \BibitemOpen
  \bibfield  {author} {\bibinfo {author} {\bibfnamefont {P.~W.}\ \bibnamefont
  {Higgs}},\ }\href {\doibase 10.1103/PhysRevLett.13.508} {\bibfield  {journal}
  {\bibinfo  {journal} {Phys. Rev. Lett.}\ }\textbf {\bibinfo {volume} {13}},\
  \bibinfo {pages} {508} (\bibinfo {year} {1964})}\BibitemShut {NoStop}%
\bibitem [{\citenamefont {Englert}\ and\ \citenamefont
  {Brout}(1964)}]{Englert:1964et}%
  \BibitemOpen
  \bibfield  {author} {\bibinfo {author} {\bibfnamefont {F.}~\bibnamefont
  {Englert}}\ and\ \bibinfo {author} {\bibfnamefont {R.}~\bibnamefont
  {Brout}},\ }\href {\doibase 10.1103/PhysRevLett.13.321} {\bibfield  {journal}
  {\bibinfo  {journal} {Phys. Rev. Lett.}\ }\textbf {\bibinfo {volume} {13}},\
  \bibinfo {pages} {321} (\bibinfo {year} {1964})}\BibitemShut {NoStop}%
\bibitem [{\citenamefont {de~Rham}\ and\ \citenamefont
  {Gabadadze}(2010)}]{deRham:2010ik}%
  \BibitemOpen
  \bibfield  {author} {\bibinfo {author} {\bibfnamefont {C.}~\bibnamefont
  {de~Rham}}\ and\ \bibinfo {author} {\bibfnamefont {G.}~\bibnamefont
  {Gabadadze}},\ }\href {\doibase 10.1103/PhysRevD.82.044020} {\bibfield
  {journal} {\bibinfo  {journal} {Phys. Rev. D}\ }\textbf {\bibinfo {volume}
  {82}},\ \bibinfo {pages} {044020} (\bibinfo {year} {2010})},\ \Eprint
  {http://arxiv.org/abs/1007.0443} {arXiv:1007.0443 [hep-th]} \BibitemShut
  {NoStop}%
\bibitem [{\citenamefont {de~Rham}\ \emph {et~al.}(2011)\citenamefont
  {de~Rham}, \citenamefont {Gabadadze},\ and\ \citenamefont
  {Tolley}}]{deRham:2010kj}%
  \BibitemOpen
  \bibfield  {author} {\bibinfo {author} {\bibfnamefont {C.}~\bibnamefont
  {de~Rham}}, \bibinfo {author} {\bibfnamefont {G.}~\bibnamefont {Gabadadze}},
  \ and\ \bibinfo {author} {\bibfnamefont {A.~J.}\ \bibnamefont {Tolley}},\
  }\href {\doibase 10.1103/PhysRevLett.106.231101} {\bibfield  {journal}
  {\bibinfo  {journal} {Phys. Rev. Lett.}\ }\textbf {\bibinfo {volume} {106}},\
  \bibinfo {pages} {231101} (\bibinfo {year} {2011})},\ \Eprint
  {http://arxiv.org/abs/1011.1232} {arXiv:1011.1232 [hep-th]} \BibitemShut
  {NoStop}%
\bibitem [{\citenamefont {Cheung}\ and\ \citenamefont
  {Remmen}(2016)}]{Cheung:2016yqr}%
  \BibitemOpen
  \bibfield  {author} {\bibinfo {author} {\bibfnamefont {C.}~\bibnamefont
  {Cheung}}\ and\ \bibinfo {author} {\bibfnamefont {G.~N.}\ \bibnamefont
  {Remmen}},\ }\href {\doibase 10.1007/JHEP04(2016)002} {\bibfield  {journal}
  {\bibinfo  {journal} {JHEP}\ }\textbf {\bibinfo {volume} {04}},\ \bibinfo
  {pages} {002} (\bibinfo {year} {2016})},\ \Eprint
  {http://arxiv.org/abs/1601.04068} {arXiv:1601.04068 [hep-th]} \BibitemShut
  {NoStop}%
\bibitem [{\citenamefont {Bellazzini}\ \emph {et~al.}(2023)\citenamefont
  {Bellazzini}, \citenamefont {Isabella}, \citenamefont {Ricossa},\ and\
  \citenamefont {Riva}}]{Bellazzini:2023nqj}%
  \BibitemOpen
  \bibfield  {author} {\bibinfo {author} {\bibfnamefont {B.}~\bibnamefont
  {Bellazzini}}, \bibinfo {author} {\bibfnamefont {G.}~\bibnamefont
  {Isabella}}, \bibinfo {author} {\bibfnamefont {S.}~\bibnamefont {Ricossa}}, \
  and\ \bibinfo {author} {\bibfnamefont {F.}~\bibnamefont {Riva}},\ }\href@noop
  {} {\  (\bibinfo {year} {2023})},\ \Eprint {http://arxiv.org/abs/2304.02550}
  {arXiv:2304.02550 [hep-th]} \BibitemShut {NoStop}%
\bibitem [{\citenamefont {Beisert}\ \emph {et~al.}(2011)\citenamefont
  {Beisert}, \citenamefont {Elvang}, \citenamefont {Freedman}, \citenamefont
  {Kiermaier}, \citenamefont {Morales},\ and\ \citenamefont
  {Stieberger}}]{Beisert:2010jx}%
  \BibitemOpen
  \bibfield  {author} {\bibinfo {author} {\bibfnamefont {N.}~\bibnamefont
  {Beisert}}, \bibinfo {author} {\bibfnamefont {H.}~\bibnamefont {Elvang}},
  \bibinfo {author} {\bibfnamefont {D.~Z.}\ \bibnamefont {Freedman}}, \bibinfo
  {author} {\bibfnamefont {M.}~\bibnamefont {Kiermaier}}, \bibinfo {author}
  {\bibfnamefont {A.}~\bibnamefont {Morales}}, \ and\ \bibinfo {author}
  {\bibfnamefont {S.}~\bibnamefont {Stieberger}},\ }\href {\doibase
  10.1016/j.physletb.2010.09.069} {\bibfield  {journal} {\bibinfo  {journal}
  {Phys. Lett. B}\ }\textbf {\bibinfo {volume} {694}},\ \bibinfo {pages} {265}
  (\bibinfo {year} {2011})},\ \Eprint {http://arxiv.org/abs/1009.1643}
  {arXiv:1009.1643 [hep-th]} \BibitemShut {NoStop}%
\bibitem [{\citenamefont {Bossard}\ \emph {et~al.}(2010)\citenamefont
  {Bossard}, \citenamefont {Hillmann},\ and\ \citenamefont
  {Nicolai}}]{Bossard:2010dq}%
  \BibitemOpen
  \bibfield  {author} {\bibinfo {author} {\bibfnamefont {G.}~\bibnamefont
  {Bossard}}, \bibinfo {author} {\bibfnamefont {C.}~\bibnamefont {Hillmann}}, \
  and\ \bibinfo {author} {\bibfnamefont {H.}~\bibnamefont {Nicolai}},\ }\href
  {\doibase 10.1007/JHEP12(2010)052} {\bibfield  {journal} {\bibinfo  {journal}
  {JHEP}\ }\textbf {\bibinfo {volume} {12}},\ \bibinfo {pages} {052} (\bibinfo
  {year} {2010})},\ \Eprint {http://arxiv.org/abs/1007.5472} {arXiv:1007.5472
  [hep-th]} \BibitemShut {NoStop}%
\bibitem [{\citenamefont {Heisenberg}\ and\ \citenamefont
  {Euler}(1936)}]{Heisenberg:1936nmg}%
  \BibitemOpen
  \bibfield  {author} {\bibinfo {author} {\bibfnamefont {W.}~\bibnamefont
  {Heisenberg}}\ and\ \bibinfo {author} {\bibfnamefont {H.}~\bibnamefont
  {Euler}},\ }\href {\doibase 10.1007/BF01343663} {\bibfield  {journal}
  {\bibinfo  {journal} {Z. Phys.}\ }\textbf {\bibinfo {volume} {98}},\ \bibinfo
  {pages} {714} (\bibinfo {year} {1936})},\ \Eprint
  {http://arxiv.org/abs/physics/0605038} {arXiv:physics/0605038} \BibitemShut
  {NoStop}%
\bibitem [{\citenamefont {Henriksson}\ \emph {et~al.}(2022)\citenamefont
  {Henriksson}, \citenamefont {McPeak}, \citenamefont {Russo},\ and\
  \citenamefont {Vichi}}]{Henriksson:2021ymi}%
  \BibitemOpen
  \bibfield  {author} {\bibinfo {author} {\bibfnamefont {J.}~\bibnamefont
  {Henriksson}}, \bibinfo {author} {\bibfnamefont {B.}~\bibnamefont {McPeak}},
  \bibinfo {author} {\bibfnamefont {F.}~\bibnamefont {Russo}}, \ and\ \bibinfo
  {author} {\bibfnamefont {A.}~\bibnamefont {Vichi}},\ }\href {\doibase
  10.1007/JHEP06(2022)158} {\bibfield  {journal} {\bibinfo  {journal} {JHEP}\
  }\textbf {\bibinfo {volume} {06}},\ \bibinfo {pages} {158} (\bibinfo {year}
  {2022})},\ \Eprint {http://arxiv.org/abs/2107.13009} {arXiv:2107.13009
  [hep-th]} \BibitemShut {NoStop}%
\bibitem [{\citenamefont {Carrillo~Gonz\'alez}\ \emph
  {et~al.}(2023)\citenamefont {Carrillo~Gonz\'alez}, \citenamefont {de~Rham},
  \citenamefont {Jaitly}, \citenamefont {Pozsgay},\ and\ \citenamefont
  {Tokareva}}]{CarrilloGonzalez:2023cbf}%
  \BibitemOpen
  \bibfield  {author} {\bibinfo {author} {\bibfnamefont {M.}~\bibnamefont
  {Carrillo~Gonz\'alez}}, \bibinfo {author} {\bibfnamefont {C.}~\bibnamefont
  {de~Rham}}, \bibinfo {author} {\bibfnamefont {S.}~\bibnamefont {Jaitly}},
  \bibinfo {author} {\bibfnamefont {V.}~\bibnamefont {Pozsgay}}, \ and\
  \bibinfo {author} {\bibfnamefont {A.}~\bibnamefont {Tokareva}},\ }\href@noop
  {} {\  (\bibinfo {year} {2023})},\ \Eprint {http://arxiv.org/abs/2307.04784}
  {arXiv:2307.04784 [hep-th]} \BibitemShut {NoStop}%
\bibitem [{\citenamefont {Polchinski}(1995)}]{Polchinski:1995mt}%
  \BibitemOpen
  \bibfield  {author} {\bibinfo {author} {\bibfnamefont {J.}~\bibnamefont
  {Polchinski}},\ }\href {\doibase 10.1103/PhysRevLett.75.4724} {\bibfield
  {journal} {\bibinfo  {journal} {Phys. Rev. Lett.}\ }\textbf {\bibinfo
  {volume} {75}},\ \bibinfo {pages} {4724} (\bibinfo {year} {1995})},\ \Eprint
  {http://arxiv.org/abs/hep-th/9510017} {arXiv:hep-th/9510017} \BibitemShut
  {NoStop}%
\bibitem [{\citenamefont {Cachazo}\ \emph {et~al.}(2015)\citenamefont
  {Cachazo}, \citenamefont {He},\ and\ \citenamefont {Yuan}}]{Cachazo:2014xea}%
  \BibitemOpen
  \bibfield  {author} {\bibinfo {author} {\bibfnamefont {F.}~\bibnamefont
  {Cachazo}}, \bibinfo {author} {\bibfnamefont {S.}~\bibnamefont {He}}, \ and\
  \bibinfo {author} {\bibfnamefont {E.~Y.}\ \bibnamefont {Yuan}},\ }\href
  {\doibase 10.1007/JHEP07(2015)149} {\bibfield  {journal} {\bibinfo  {journal}
  {JHEP}\ }\textbf {\bibinfo {volume} {07}},\ \bibinfo {pages} {149} (\bibinfo
  {year} {2015})},\ \Eprint {http://arxiv.org/abs/1412.3479} {arXiv:1412.3479
  [hep-th]} \BibitemShut {NoStop}%
\bibitem [{\citenamefont {Cheung}\ \emph
  {et~al.}(2018{\natexlab{a}})\citenamefont {Cheung}, \citenamefont {Kampf},
  \citenamefont {Novotny}, \citenamefont {Shen}, \citenamefont {Trnka},\ and\
  \citenamefont {Wen}}]{Cheung:2018oki}%
  \BibitemOpen
  \bibfield  {author} {\bibinfo {author} {\bibfnamefont {C.}~\bibnamefont
  {Cheung}}, \bibinfo {author} {\bibfnamefont {K.}~\bibnamefont {Kampf}},
  \bibinfo {author} {\bibfnamefont {J.}~\bibnamefont {Novotny}}, \bibinfo
  {author} {\bibfnamefont {C.-H.}\ \bibnamefont {Shen}}, \bibinfo {author}
  {\bibfnamefont {J.}~\bibnamefont {Trnka}}, \ and\ \bibinfo {author}
  {\bibfnamefont {C.}~\bibnamefont {Wen}},\ }\href {\doibase
  10.1103/PhysRevLett.120.261602} {\bibfield  {journal} {\bibinfo  {journal}
  {Phys. Rev. Lett.}\ }\textbf {\bibinfo {volume} {120}},\ \bibinfo {pages}
  {261602} (\bibinfo {year} {2018}{\natexlab{a}})},\ \Eprint
  {http://arxiv.org/abs/1801.01496} {arXiv:1801.01496 [hep-th]} \BibitemShut
  {NoStop}%
\bibitem [{\citenamefont {Bern}\ \emph {et~al.}(2010)\citenamefont {Bern},
  \citenamefont {Carrasco},\ and\ \citenamefont {Johansson}}]{Bern:2010ue}%
  \BibitemOpen
  \bibfield  {author} {\bibinfo {author} {\bibfnamefont {Z.}~\bibnamefont
  {Bern}}, \bibinfo {author} {\bibfnamefont {J.~J.~M.}\ \bibnamefont
  {Carrasco}}, \ and\ \bibinfo {author} {\bibfnamefont {H.}~\bibnamefont
  {Johansson}},\ }\href {\doibase 10.1103/PhysRevLett.105.061602} {\bibfield
  {journal} {\bibinfo  {journal} {Phys. Rev. Lett.}\ }\textbf {\bibinfo
  {volume} {105}},\ \bibinfo {pages} {061602} (\bibinfo {year} {2010})},\
  \Eprint {http://arxiv.org/abs/1004.0476} {arXiv:1004.0476 [hep-th]}
  \BibitemShut {NoStop}%
\bibitem [{\citenamefont {de~Neeling}\ \emph {et~al.}(2022)\citenamefont
  {de~Neeling}, \citenamefont {Roest},\ and\ \citenamefont
  {Veldmeijer}}]{deNeeling:2022tsu}%
  \BibitemOpen
  \bibfield  {author} {\bibinfo {author} {\bibfnamefont {D.}~\bibnamefont
  {de~Neeling}}, \bibinfo {author} {\bibfnamefont {D.}~\bibnamefont {Roest}}, \
  and\ \bibinfo {author} {\bibfnamefont {S.}~\bibnamefont {Veldmeijer}},\
  }\href {\doibase 10.1007/JHEP10(2022)066} {\bibfield  {journal} {\bibinfo
  {journal} {JHEP}\ }\textbf {\bibinfo {volume} {10}},\ \bibinfo {pages} {066}
  (\bibinfo {year} {2022})},\ \Eprint {http://arxiv.org/abs/2204.11629}
  {arXiv:2204.11629 [hep-th]} \BibitemShut {NoStop}%
\bibitem [{\citenamefont {Cheung}\ \emph
  {et~al.}(2018{\natexlab{b}})\citenamefont {Cheung}, \citenamefont {Shen},\
  and\ \citenamefont {Wen}}]{Cheung:2017ems}%
  \BibitemOpen
  \bibfield  {author} {\bibinfo {author} {\bibfnamefont {C.}~\bibnamefont
  {Cheung}}, \bibinfo {author} {\bibfnamefont {C.-H.}\ \bibnamefont {Shen}}, \
  and\ \bibinfo {author} {\bibfnamefont {C.}~\bibnamefont {Wen}},\ }\href
  {\doibase 10.1007/JHEP02(2018)095} {\bibfield  {journal} {\bibinfo  {journal}
  {JHEP}\ }\textbf {\bibinfo {volume} {02}},\ \bibinfo {pages} {095} (\bibinfo
  {year} {2018}{\natexlab{b}})},\ \Eprint {http://arxiv.org/abs/1705.03025}
  {arXiv:1705.03025 [hep-th]} \BibitemShut {NoStop}%
\bibitem [{\citenamefont {Tolley}(2020)}]{Tolley:2019nmm}%
  \BibitemOpen
  \bibfield  {author} {\bibinfo {author} {\bibfnamefont {A.~J.}\ \bibnamefont
  {Tolley}},\ }\href {\doibase 10.1007/JHEP06(2020)050} {\bibfield  {journal}
  {\bibinfo  {journal} {JHEP}\ }\textbf {\bibinfo {volume} {06}},\ \bibinfo
  {pages} {050} (\bibinfo {year} {2020})},\ \Eprint
  {http://arxiv.org/abs/1911.06142} {arXiv:1911.06142 [hep-th]} \BibitemShut
  {NoStop}%
\bibitem [{\citenamefont {Cavagli\`a}\ \emph {et~al.}(2016)\citenamefont
  {Cavagli\`a}, \citenamefont {Negro}, \citenamefont {Sz\'ecs\'enyi},\ and\
  \citenamefont {Tateo}}]{Cavaglia:2016oda}%
  \BibitemOpen
  \bibfield  {author} {\bibinfo {author} {\bibfnamefont {A.}~\bibnamefont
  {Cavagli\`a}}, \bibinfo {author} {\bibfnamefont {S.}~\bibnamefont {Negro}},
  \bibinfo {author} {\bibfnamefont {I.~M.}\ \bibnamefont {Sz\'ecs\'enyi}}, \
  and\ \bibinfo {author} {\bibfnamefont {R.}~\bibnamefont {Tateo}},\ }\href
  {\doibase 10.1007/JHEP10(2016)112} {\bibfield  {journal} {\bibinfo  {journal}
  {JHEP}\ }\textbf {\bibinfo {volume} {10}},\ \bibinfo {pages} {112} (\bibinfo
  {year} {2016})},\ \Eprint {http://arxiv.org/abs/1608.05534} {arXiv:1608.05534
  [hep-th]} \BibitemShut {NoStop}%
\bibitem [{\citenamefont {Conti}\ \emph {et~al.}(2018)\citenamefont {Conti},
  \citenamefont {Iannella}, \citenamefont {Negro},\ and\ \citenamefont
  {Tateo}}]{Conti:2018jho}%
  \BibitemOpen
  \bibfield  {author} {\bibinfo {author} {\bibfnamefont {R.}~\bibnamefont
  {Conti}}, \bibinfo {author} {\bibfnamefont {L.}~\bibnamefont {Iannella}},
  \bibinfo {author} {\bibfnamefont {S.}~\bibnamefont {Negro}}, \ and\ \bibinfo
  {author} {\bibfnamefont {R.}~\bibnamefont {Tateo}},\ }\href {\doibase
  10.1007/JHEP11(2018)007} {\bibfield  {journal} {\bibinfo  {journal} {JHEP}\
  }\textbf {\bibinfo {volume} {11}},\ \bibinfo {pages} {007} (\bibinfo {year}
  {2018})},\ \Eprint {http://arxiv.org/abs/1806.11515} {arXiv:1806.11515
  [hep-th]} \BibitemShut {NoStop}%
\bibitem [{\citenamefont {Babaei-Aghbolagh}\ \emph {et~al.}(2022)\citenamefont
  {Babaei-Aghbolagh}, \citenamefont {Velni}, \citenamefont {Yekta},\ and\
  \citenamefont {Mohammadzadeh}}]{Babaei-Aghbolagh:2022uij}%
  \BibitemOpen
  \bibfield  {author} {\bibinfo {author} {\bibfnamefont {H.}~\bibnamefont
  {Babaei-Aghbolagh}}, \bibinfo {author} {\bibfnamefont {K.~B.}\ \bibnamefont
  {Velni}}, \bibinfo {author} {\bibfnamefont {D.~M.}\ \bibnamefont {Yekta}}, \
  and\ \bibinfo {author} {\bibfnamefont {H.}~\bibnamefont {Mohammadzadeh}},\
  }\href {\doibase 10.1016/j.physletb.2022.137079} {\bibfield  {journal}
  {\bibinfo  {journal} {Phys. Lett. B}\ }\textbf {\bibinfo {volume} {829}},\
  \bibinfo {pages} {137079} (\bibinfo {year} {2022})},\ \Eprint
  {http://arxiv.org/abs/2202.11156} {arXiv:2202.11156 [hep-th]} \BibitemShut
  {NoStop}%
\bibitem [{\citenamefont {Ferko}\ \emph {et~al.}(2022)\citenamefont {Ferko},
  \citenamefont {Smith},\ and\ \citenamefont
  {Tartaglino-Mazzucchelli}}]{Ferko:2022iru}%
  \BibitemOpen
  \bibfield  {author} {\bibinfo {author} {\bibfnamefont {C.}~\bibnamefont
  {Ferko}}, \bibinfo {author} {\bibfnamefont {L.}~\bibnamefont {Smith}}, \ and\
  \bibinfo {author} {\bibfnamefont {G.}~\bibnamefont
  {Tartaglino-Mazzucchelli}},\ }\href {\doibase 10.21468/SciPostPhys.13.2.012}
  {\bibfield  {journal} {\bibinfo  {journal} {SciPost Phys.}\ }\textbf
  {\bibinfo {volume} {13}},\ \bibinfo {pages} {012} (\bibinfo {year} {2022})},\
  \Eprint {http://arxiv.org/abs/2203.01085} {arXiv:2203.01085 [hep-th]}
  \BibitemShut {NoStop}%
\bibitem [{\citenamefont {Conti}\ \emph {et~al.}(2022)\citenamefont {Conti},
  \citenamefont {Romano},\ and\ \citenamefont {Tateo}}]{Conti:2022egv}%
  \BibitemOpen
  \bibfield  {author} {\bibinfo {author} {\bibfnamefont {R.}~\bibnamefont
  {Conti}}, \bibinfo {author} {\bibfnamefont {J.}~\bibnamefont {Romano}}, \
  and\ \bibinfo {author} {\bibfnamefont {R.}~\bibnamefont {Tateo}},\ }\href
  {\doibase 10.1007/JHEP09(2022)085} {\bibfield  {journal} {\bibinfo  {journal}
  {JHEP}\ }\textbf {\bibinfo {volume} {09}},\ \bibinfo {pages} {085} (\bibinfo
  {year} {2022})},\ \Eprint {http://arxiv.org/abs/2206.03415} {arXiv:2206.03415
  [hep-th]} \BibitemShut {NoStop}%
\bibitem [{\citenamefont {Hinterbichler}(2012)}]{Hinterbichler:2011tt}%
  \BibitemOpen
  \bibfield  {author} {\bibinfo {author} {\bibfnamefont {K.}~\bibnamefont
  {Hinterbichler}},\ }\href {\doibase 10.1103/RevModPhys.84.671} {\bibfield
  {journal} {\bibinfo  {journal} {Rev. Mod. Phys.}\ }\textbf {\bibinfo {volume}
  {84}},\ \bibinfo {pages} {671} (\bibinfo {year} {2012})},\ \Eprint
  {http://arxiv.org/abs/1105.3735} {arXiv:1105.3735 [hep-th]} \BibitemShut
  {NoStop}%
\bibitem [{\citenamefont
  {Bialynicki-Birula}(1983)}]{Bialynicki-Birula:1984daz}%
  \BibitemOpen
  \bibfield  {author} {\bibinfo {author} {\bibfnamefont {I.}~\bibnamefont
  {Bialynicki-Birula}},\ }in\ \href@noop {} {\emph {\bibinfo {booktitle} {{B.
  Jancewicz and J. Lukierski eds/, Quantum Theory of Particles and Fields,
  World Scientific}}}}\ (\bibinfo {year} {1983})\ pp.\ \bibinfo {pages}
  {31--48}\BibitemShut {NoStop}%
\bibitem [{\citenamefont {Sorokin}(2022)}]{Sorokin:2021tge}%
  \BibitemOpen
  \bibfield  {author} {\bibinfo {author} {\bibfnamefont {D.~P.}\ \bibnamefont
  {Sorokin}},\ }\href {\doibase 10.1002/prop.202200092} {\bibfield  {journal}
  {\bibinfo  {journal} {Fortsch. Phys.}\ }\textbf {\bibinfo {volume} {70}},\
  \bibinfo {pages} {2200092} (\bibinfo {year} {2022})},\ \Eprint
  {http://arxiv.org/abs/2112.12118} {arXiv:2112.12118 [hep-th]} \BibitemShut
  {NoStop}%
\bibitem [{\citenamefont {Gibbons}\ and\ \citenamefont
  {Rasheed}(1995)}]{Gibbons:1995cv}%
  \BibitemOpen
  \bibfield  {author} {\bibinfo {author} {\bibfnamefont {G.~W.}\ \bibnamefont
  {Gibbons}}\ and\ \bibinfo {author} {\bibfnamefont {D.~A.}\ \bibnamefont
  {Rasheed}},\ }\href {\doibase 10.1016/0550-3213(95)00409-L} {\bibfield
  {journal} {\bibinfo  {journal} {Nucl. Phys. B}\ }\textbf {\bibinfo {volume}
  {454}},\ \bibinfo {pages} {185} (\bibinfo {year} {1995})},\ \Eprint
  {http://arxiv.org/abs/hep-th/9506035} {arXiv:hep-th/9506035} \BibitemShut
  {NoStop}%
\bibitem [{\citenamefont {Gaillard}\ and\ \citenamefont
  {Zumino}(1997)}]{Gaillard:1997rt}%
  \BibitemOpen
  \bibfield  {author} {\bibinfo {author} {\bibfnamefont {M.~K.}\ \bibnamefont
  {Gaillard}}\ and\ \bibinfo {author} {\bibfnamefont {B.}~\bibnamefont
  {Zumino}},\ }in\ \href@noop {} {\emph {\bibinfo {booktitle} {{A Newton
  Institute Euroconference on Duality and Supersymmetric Theories}}}}\
  (\bibinfo {year} {1997})\ pp.\ \bibinfo {pages} {33--48},\ \Eprint
  {http://arxiv.org/abs/hep-th/9712103} {arXiv:hep-th/9712103} \BibitemShut
  {NoStop}%
\bibitem [{\citenamefont {Bandos}\ \emph {et~al.}(2020)\citenamefont {Bandos},
  \citenamefont {Lechner}, \citenamefont {Sorokin},\ and\ \citenamefont
  {Townsend}}]{Bandos:2020jsw}%
  \BibitemOpen
  \bibfield  {author} {\bibinfo {author} {\bibfnamefont {I.}~\bibnamefont
  {Bandos}}, \bibinfo {author} {\bibfnamefont {K.}~\bibnamefont {Lechner}},
  \bibinfo {author} {\bibfnamefont {D.}~\bibnamefont {Sorokin}}, \ and\
  \bibinfo {author} {\bibfnamefont {P.~K.}\ \bibnamefont {Townsend}},\ }\href
  {\doibase 10.1103/PhysRevD.102.121703} {\bibfield  {journal} {\bibinfo
  {journal} {Phys. Rev. D}\ }\textbf {\bibinfo {volume} {102}},\ \bibinfo
  {pages} {121703} (\bibinfo {year} {2020})},\ \Eprint
  {http://arxiv.org/abs/2007.09092} {arXiv:2007.09092 [hep-th]} \BibitemShut
  {NoStop}%
\bibitem [{\citenamefont {Born}\ and\ \citenamefont
  {Infeld}(1934)}]{Born:1934gh}%
  \BibitemOpen
  \bibfield  {author} {\bibinfo {author} {\bibfnamefont {M.}~\bibnamefont
  {Born}}\ and\ \bibinfo {author} {\bibfnamefont {L.}~\bibnamefont {Infeld}},\
  }\href {\doibase 10.1098/rspa.1934.0059} {\bibfield  {journal} {\bibinfo
  {journal} {Proc. Roy. Soc. Lond. A}\ }\textbf {\bibinfo {volume} {144}},\
  \bibinfo {pages} {425} (\bibinfo {year} {1934})}\BibitemShut {NoStop}%
\bibitem [{\citenamefont {Hatsuda}\ \emph {et~al.}(1999)\citenamefont
  {Hatsuda}, \citenamefont {Kamimura},\ and\ \citenamefont
  {Sekiya}}]{Hatsuda:1999ys}%
  \BibitemOpen
  \bibfield  {author} {\bibinfo {author} {\bibfnamefont {M.}~\bibnamefont
  {Hatsuda}}, \bibinfo {author} {\bibfnamefont {K.}~\bibnamefont {Kamimura}}, \
  and\ \bibinfo {author} {\bibfnamefont {S.}~\bibnamefont {Sekiya}},\ }\href
  {\doibase 10.1016/S0550-3213(99)00509-X} {\bibfield  {journal} {\bibinfo
  {journal} {Nucl. Phys. B}\ }\textbf {\bibinfo {volume} {561}},\ \bibinfo
  {pages} {341} (\bibinfo {year} {1999})},\ \Eprint
  {http://arxiv.org/abs/hep-th/9906103} {arXiv:hep-th/9906103} \BibitemShut
  {NoStop}%
\bibitem [{\citenamefont {Kuzenko}\ and\ \citenamefont
  {Theisen}(2000)}]{Kuzenko:2000tg}%
  \BibitemOpen
  \bibfield  {author} {\bibinfo {author} {\bibfnamefont {S.~M.}\ \bibnamefont
  {Kuzenko}}\ and\ \bibinfo {author} {\bibfnamefont {S.}~\bibnamefont
  {Theisen}},\ }\href {\doibase 10.1088/1126-6708/2000/03/034} {\bibfield
  {journal} {\bibinfo  {journal} {JHEP}\ }\textbf {\bibinfo {volume} {03}},\
  \bibinfo {pages} {034} (\bibinfo {year} {2000})},\ \Eprint
  {http://arxiv.org/abs/hep-th/0001068} {arXiv:hep-th/0001068} \BibitemShut
  {NoStop}%
\bibitem [{\citenamefont {Ivanov}\ \emph {et~al.}(2013)\citenamefont {Ivanov},
  \citenamefont {Lechtenfeld},\ and\ \citenamefont {Zupnik}}]{Ivanov:2013ppa}%
  \BibitemOpen
  \bibfield  {author} {\bibinfo {author} {\bibfnamefont {E.}~\bibnamefont
  {Ivanov}}, \bibinfo {author} {\bibfnamefont {O.}~\bibnamefont {Lechtenfeld}},
  \ and\ \bibinfo {author} {\bibfnamefont {B.}~\bibnamefont {Zupnik}},\ }\href
  {\doibase 10.1007/JHEP05(2013)133} {\bibfield  {journal} {\bibinfo  {journal}
  {JHEP}\ }\textbf {\bibinfo {volume} {05}},\ \bibinfo {pages} {133} (\bibinfo
  {year} {2013})},\ \Eprint {http://arxiv.org/abs/1303.5962} {arXiv:1303.5962
  [hep-th]} \BibitemShut {NoStop}%
\bibitem [{\citenamefont {Carrasco}\ \emph {et~al.}(2012)\citenamefont
  {Carrasco}, \citenamefont {Kallosh},\ and\ \citenamefont
  {Roiban}}]{Carrasco:2011jv}%
  \BibitemOpen
  \bibfield  {author} {\bibinfo {author} {\bibfnamefont {J.~J.~M.}\
  \bibnamefont {Carrasco}}, \bibinfo {author} {\bibfnamefont {R.}~\bibnamefont
  {Kallosh}}, \ and\ \bibinfo {author} {\bibfnamefont {R.}~\bibnamefont
  {Roiban}},\ }\href {\doibase 10.1103/PhysRevD.85.025007} {\bibfield
  {journal} {\bibinfo  {journal} {Phys. Rev. D}\ }\textbf {\bibinfo {volume}
  {85}},\ \bibinfo {pages} {025007} (\bibinfo {year} {2012})},\ \Eprint
  {http://arxiv.org/abs/1108.4390} {arXiv:1108.4390 [hep-th]} \BibitemShut
  {NoStop}%
\bibitem [{\citenamefont {Ivanov}\ and\ \citenamefont
  {Zupnik}(2002)}]{Ivanov:2002ab}%
  \BibitemOpen
  \bibfield  {author} {\bibinfo {author} {\bibfnamefont {E.~A.}\ \bibnamefont
  {Ivanov}}\ and\ \bibinfo {author} {\bibfnamefont {B.~M.}\ \bibnamefont
  {Zupnik}},\ }in\ \href@noop {} {\emph {\bibinfo {booktitle} {{4th
  International Workshop on Supersymmetry and Quantum Symmetries}: {16th Max
  Born Symposium}}}}\ (\bibinfo {year} {2002})\ pp.\ \bibinfo {pages}
  {235--250},\ \Eprint {http://arxiv.org/abs/hep-th/0202203}
  {arXiv:hep-th/0202203} \BibitemShut {NoStop}%
\bibitem [{\citenamefont {Boulware}\ and\ \citenamefont
  {Deser}(1972)}]{Boulware:1972yco}%
  \BibitemOpen
  \bibfield  {author} {\bibinfo {author} {\bibfnamefont {D.~G.}\ \bibnamefont
  {Boulware}}\ and\ \bibinfo {author} {\bibfnamefont {S.}~\bibnamefont
  {Deser}},\ }\href {\doibase 10.1103/PhysRevD.6.3368} {\bibfield  {journal}
  {\bibinfo  {journal} {Phys. Rev. D}\ }\textbf {\bibinfo {volume} {6}},\
  \bibinfo {pages} {3368} (\bibinfo {year} {1972})}\BibitemShut {NoStop}%
\bibitem [{\citenamefont {Fierz}\ and\ \citenamefont
  {Pauli}(1939)}]{Fierz:1939ix}%
  \BibitemOpen
  \bibfield  {author} {\bibinfo {author} {\bibfnamefont {M.}~\bibnamefont
  {Fierz}}\ and\ \bibinfo {author} {\bibfnamefont {W.}~\bibnamefont {Pauli}},\
  }\href {\doibase 10.1098/rspa.1939.0140} {\bibfield  {journal} {\bibinfo
  {journal} {Proc. Roy. Soc. Lond. A}\ }\textbf {\bibinfo {volume} {173}},\
  \bibinfo {pages} {211} (\bibinfo {year} {1939})}\BibitemShut {NoStop}%
\bibitem [{\citenamefont {Hinterbichler}\ and\ \citenamefont
  {Joyce}(2015)}]{Hinterbichler:2015pqa}%
  \BibitemOpen
  \bibfield  {author} {\bibinfo {author} {\bibfnamefont {K.}~\bibnamefont
  {Hinterbichler}}\ and\ \bibinfo {author} {\bibfnamefont {A.}~\bibnamefont
  {Joyce}},\ }\href {\doibase 10.1103/PhysRevD.92.023503} {\bibfield  {journal}
  {\bibinfo  {journal} {Phys. Rev. D}\ }\textbf {\bibinfo {volume} {92}},\
  \bibinfo {pages} {023503} (\bibinfo {year} {2015})},\ \Eprint
  {http://arxiv.org/abs/1501.07600} {arXiv:1501.07600 [hep-th]} \BibitemShut
  {NoStop}%
\bibitem [{\citenamefont {Sen}(2020)}]{Sen:2019qit}%
  \BibitemOpen
  \bibfield  {author} {\bibinfo {author} {\bibfnamefont {A.}~\bibnamefont
  {Sen}},\ }\href {\doibase 10.1088/1751-8121/ab5423} {\bibfield  {journal}
  {\bibinfo  {journal} {J. Phys. A}\ }\textbf {\bibinfo {volume} {53}},\
  \bibinfo {pages} {084002} (\bibinfo {year} {2020})},\ \Eprint
  {http://arxiv.org/abs/1903.12196} {arXiv:1903.12196 [hep-th]} \BibitemShut
  {NoStop}%
\bibitem [{\citenamefont {Hull}(2023)}]{Hull:2023dgp}%
  \BibitemOpen
  \bibfield  {author} {\bibinfo {author} {\bibfnamefont {C.~M.}\ \bibnamefont
  {Hull}},\ }\href@noop {} {\  (\bibinfo {year} {2023})},\ \Eprint
  {http://arxiv.org/abs/2307.04748} {arXiv:2307.04748 [hep-th]} \BibitemShut
  {NoStop}%
\bibitem [{\citenamefont {Bandos}\ \emph {et~al.}(2021)\citenamefont {Bandos},
  \citenamefont {Lechner}, \citenamefont {Sorokin},\ and\ \citenamefont
  {Townsend}}]{Bandos:2020hgy}%
  \BibitemOpen
  \bibfield  {author} {\bibinfo {author} {\bibfnamefont {I.}~\bibnamefont
  {Bandos}}, \bibinfo {author} {\bibfnamefont {K.}~\bibnamefont {Lechner}},
  \bibinfo {author} {\bibfnamefont {D.}~\bibnamefont {Sorokin}}, \ and\
  \bibinfo {author} {\bibfnamefont {P.~K.}\ \bibnamefont {Townsend}},\ }\href
  {\doibase 10.1007/JHEP03(2021)022} {\bibfield  {journal} {\bibinfo  {journal}
  {JHEP}\ }\textbf {\bibinfo {volume} {03}},\ \bibinfo {pages} {022} (\bibinfo
  {year} {2021})},\ \Eprint {http://arxiv.org/abs/2012.09286} {arXiv:2012.09286
  [hep-th]} \BibitemShut {NoStop}%
\bibitem [{\citenamefont {Cachazo}\ \emph {et~al.}(2014)\citenamefont
  {Cachazo}, \citenamefont {He},\ and\ \citenamefont {Yuan}}]{Cachazo:2013hca}%
  \BibitemOpen
  \bibfield  {author} {\bibinfo {author} {\bibfnamefont {F.}~\bibnamefont
  {Cachazo}}, \bibinfo {author} {\bibfnamefont {S.}~\bibnamefont {He}}, \ and\
  \bibinfo {author} {\bibfnamefont {E.~Y.}\ \bibnamefont {Yuan}},\ }\href
  {\doibase 10.1103/PhysRevLett.113.171601} {\bibfield  {journal} {\bibinfo
  {journal} {Phys. Rev. Lett.}\ }\textbf {\bibinfo {volume} {113}},\ \bibinfo
  {pages} {171601} (\bibinfo {year} {2014})},\ \Eprint
  {http://arxiv.org/abs/1307.2199} {arXiv:1307.2199 [hep-th]} \BibitemShut
  {NoStop}%
\bibitem [{\citenamefont {Cheung}\ \emph {et~al.}(2015)\citenamefont {Cheung},
  \citenamefont {Kampf}, \citenamefont {Novotny},\ and\ \citenamefont
  {Trnka}}]{Cheung:2014dqa}%
  \BibitemOpen
  \bibfield  {author} {\bibinfo {author} {\bibfnamefont {C.}~\bibnamefont
  {Cheung}}, \bibinfo {author} {\bibfnamefont {K.}~\bibnamefont {Kampf}},
  \bibinfo {author} {\bibfnamefont {J.}~\bibnamefont {Novotny}}, \ and\
  \bibinfo {author} {\bibfnamefont {J.}~\bibnamefont {Trnka}},\ }\href
  {\doibase 10.1103/PhysRevLett.114.221602} {\bibfield  {journal} {\bibinfo
  {journal} {Phys. Rev. Lett.}\ }\textbf {\bibinfo {volume} {114}},\ \bibinfo
  {pages} {221602} (\bibinfo {year} {2015})},\ \Eprint
  {http://arxiv.org/abs/1412.4095} {arXiv:1412.4095 [hep-th]} \BibitemShut
  {NoStop}%
\bibitem [{\citenamefont {Volkov}\ and\ \citenamefont
  {Akulov}(1973)}]{Volkov:1973ix}%
  \BibitemOpen
  \bibfield  {author} {\bibinfo {author} {\bibfnamefont {D.~V.}\ \bibnamefont
  {Volkov}}\ and\ \bibinfo {author} {\bibfnamefont {V.~P.}\ \bibnamefont
  {Akulov}},\ }\href {\doibase 10.1016/0370-2693(73)90490-5} {\bibfield
  {journal} {\bibinfo  {journal} {Phys. Lett. B}\ }\textbf {\bibinfo {volume}
  {46}},\ \bibinfo {pages} {109} (\bibinfo {year} {1973})}\BibitemShut
  {NoStop}%
\bibitem [{\citenamefont {Mart\'in-Garc\'ia}()}]{xAct}%
  \BibitemOpen
  \bibfield  {author} {\bibinfo {author} {\bibfnamefont {J.~M.}\ \bibnamefont
  {Mart\'in-Garc\'ia}},\ }\href@noop {} {\ }\bibinfo {note}
  {\url{http://www.xact.es/}}\BibitemShut {NoStop}%
\bibitem [{\citenamefont {Nutma}(2014)}]{Nutma:2013zea}%
  \BibitemOpen
  \bibfield  {author} {\bibinfo {author} {\bibfnamefont {T.}~\bibnamefont
  {Nutma}},\ }\href {\doibase 10.1016/j.cpc.2014.02.006} {\bibfield  {journal}
  {\bibinfo  {journal} {Comput. Phys. Commun.}\ }\textbf {\bibinfo {volume}
  {185}},\ \bibinfo {pages} {1719} (\bibinfo {year} {2014})},\ \Eprint
  {http://arxiv.org/abs/1308.3493} {arXiv:1308.3493 [cs.SC]} \BibitemShut
  {NoStop}%
\end{thebibliography}%

\end{document}